\documentclass[showpacs,notitlepage,prd,aps,longbibliography,twocolumn]{revtex4-1}
\usepackage{graphicx}\usepackage{amsmath}\usepackage{amssymb}\usepackage{slashed}

\newcommand{\be}{\begin{equation}}\newcommand{\ee}{\end{equation}}
\newcommand{\bea}{\begin{eqnarray}}\newcommand{\eea}{\end{eqnarray}}
\newcommand{\nn}{\nonumber}
\newcommand{\pa}{\partial}
\newcommand{\ben}{\begin{enumerate}}\newcommand{\een}{\end{enumerate}}

\newcommand{\om}{\omega}\newcommand{\Om}{\Omega}

\renewcommand{\kappa}{\varkappa}
\newcommand{\q}{\mathbf{q}}
\newcommand{\x}{\mathbf{x}}
\renewcommand{\a}{\mathbf{a}}\renewcommand{\c}{\mathbf{c}}
\renewcommand{\k}{\mathbf{k}}

\newcommand{\n}{\mathbf{n}}\newcommand{\m}{\mathbf{m}}
\newcommand{\Nn}{\mathbf{N}}\newcommand{\Mm}{\mathbf{M}}

\newcommand{\tfi}{\tilde{\phi}}
\newcommand{\Tr}{{\rm Tr}}
\newcommand{\M}{{\cal M}}\newcommand{\N}{{\cal N}}
\newcommand{\Ref}[1]{(\ref{#1})}

\newcommand{\book}{\cite{BKMM}}
\usepackage{color}

\begin{document}

\title{Casimir effect for Dirac lattices}

\author{M. Bordag}
\affiliation{Institut f\"ur Theoretische Physik, Universit\"at Leipzig, Germany}
\email{bordag@uni-leipzig.de}

\author{I.G. Pirozhenko}
\affiliation{Bogoliubov Laboratory of Theoretical Physics, JINR, Dubna, Russia\\Dubna State University, Dubna, Russia}
\email{pirozhen@theor.jinr.ru}

%
%
\begin{abstract}
We consider polarizable sheets, which recently received some attention, especially in context of the dispersion interaction of thin sheets like graphene. These sheets are modeled by a collection of delta function potentials and resemble ‘zero range potentials’, known in quantum mechanics. We develop a theoretical description and apply the so-called ‘TGTG’-formula to calculate the interaction of two such lattices. Thereby we make use of the formulation of the scattering of waves off such sheets provided earlier. We consider all limiting cases, providing link to earlier results. Also, we discuss the relation to the pairwise summation method.

\end{abstract}
%
\maketitle
\section{Introduction}
This paper is a contribution to the discussion of van der Waals and Casimir forces between surfaces. Last years, much attention was paid to the interaction between slabs of finite, and especially small,  thickness. This triggered by a growing interest to two dimensional structures (sheets) like 2d electron gas, monoatomically layers and graphene, as well as to the interaction between them. As discussed in \cite{bart13-15-063028}, the situation with sheets having only in-plane polarizability is relatively clear. Here one can formulate a hydrodynamic plasma model and calculate the quantities of interest \cite{BV}. The same holds for graphene described by the Dirac equation model for the $\pi$-electrons \cite{bord09-80-245406}, which are responsible for the interaction with the electromagnetic field.

The situation with perpendicular polarizability is more complicated. While in \cite{para12-89-085021} no response to the electromagnetic field was found, in \cite{bart15-354-534} it was shown that such response takes place. In that paper the sheet was modelled as a lattice of harmonic oscillators (dipoles), vibrating in direction perpendicular to the sheet, and the limit of zero spacing of this lattice was investigated. This was extended in \cite{bord14-89-125015} to a  sheet, continuous from the very beginning, having parallel or perpendicular polarizabilities. Later, a similar setup was considered using point dipoles \cite{bord15-91-065027}. These can be represented  by Dirac delta functions forming a so-called 'Dirac lattice'. In fact, there represent point scatterers.

Such lattices, taken alone, have a well known internal dynamics, the simplest case being the Kronig-Penney model ('Dirac comb') \cite{kron31-130-499}. In quantum mechanics their use is known as 'method of zero-range potential' \cite{demk88}. As well known, in more than one dimension a Hamilton operator with a delta function potential is not self adjoint \cite{bere61-2-372}. In electrodynamics, the self energy of a delta function potential is singular and one needs a renormalization. In terms of quantum field theory this setup was considered in \cite{jack95}.

Recently, a sheet of delta function potentials was used in \cite{bord15-91-065027} to model a polarizable sheet. For instance, scattering off such sheet was investigated and subsequently the transition to a continuous sheet as well.

In the present note we consider a setup of two such sheets, hold in parallel at some separation. This is a typical situation for the Casimir effect.
We use the well known scattering approach, also called 'TGTG'-formula, which can be considered also as a generalization of the Lifshitz formula. Making use of the translational invariance of the lattice, the $T$-matrices can be calculated in momentum representation; for a one dimensional lattice even explicitly. Further, we allow for a displacement of one lattice with respect to the other. Also we study the limiting cases of small and large separation, resp. of large and small lattice spacing. For large separation, the limi\-ting case corresponds to the interaction of two sheets carrying delta function potential, and for small separation the limiting case corresponds to the interaction of two points carrying delta function potential.  In this paper we consider only a scalar field. Extensions to the electromagnetic field should be quite straight forward.
It is also an intention of this paper to provide a framework for calculating dispersion forces in realistic, experimentally relevant situations, where the atomistic structure must be accounted for.

The paper is organized as follows. In the next section we provide the necessary formulas to calculate the vacuum energy for two and one dimensional lattices. In the third section we derive all limiting cases and discuss their interrelations. In the fourth section we compare the results with pairwise summation. Some conclusions and an appendix complete the paper.

\maketitle
\section{Vacuum energy for point scatterers}
We consider two types of lattices. First, two dimensional lattices, A and B, given by the lattice vectors
\be \vec{a}^{\rm \,A}_{\n}=\left(\begin{array}{c}\a_\n+\c \\ b\end{array}\right),
    \quad   \vec{a}^{\rm \,B}_{\n}=\left(\begin{array}{c}\a_\n  \\0\end{array}\right),
\label{2.1}\ee
For these, we use the following notations. A three dimensional vector is denoted by an arrow, a two dimensional vector parallel to the $(x,y)$-plane is denoted in bold, for instance $\vec{x}=\left(\begin{array}{c}\x \\ x_3\end{array}\right)$. The lattice sites are given by
$\a_\n=a\left(\begin{array}{c} n_1 \\ n_2\end{array}\right)$,
where $n_1$ and $n_2$ are integers, $a$ is the lattice spacing; the lattice is rectangular.
The lattice B is in the $(x,y)$-plane and the lattice A is parallel to B on a separation $b$ and shifted within that plane by the displacement vector $\c$.

Second we consider one dimensional lattices (chains), A and B, given by the lattice vectors
\be \vec{a}^{\rm \,A}_{n}=\left(        \begin{array}{c}a n+c \\0\\ b\end{array}\right),
    \quad   \vec{a}^{\rm \,B}_{n}=\left(\begin{array}{c}an \\0 \\0\end{array}\right),
\label{2.1a}\ee
where $n$ is a (single) integer. These chains are on the $x$-axis (B) resp. parallel to it (A). Both are in the plane $y=0$. Their separation is $b$ as above, the displacement $c$ is a number.

With these notations, we consider a scalar field $\phi(\vec{x})$, whose wave equation after Fourier transform in time is
\be \left(-\om^2-\Delta+g\sum_{\n}\left(
    \delta(\vec{x}-\vec{a}^{\rm \,A}_{\n})+\delta(\vec{x}-\vec{a}^{\rm \,B}_{\n})\right)
              \right)\phi(\vec{x})=0,
\label{2.2}\ee
where $\vec{a}^{\rm \,A}_{\n}$ and $\vec{a}^{\rm \,B}_{\n}$ are given by \Ref{2.1}. In this equation, the delta functions are three dimensional, hence the coupling $g$ has the dimension of length. As mentioned in the Introduction, one should remember that the above equation is not well defined and $g$ must be viewed as a bare coupling which needs to undergo a renormalization.

The vacuum energy of the interaction of two lattices is given by the mentioned 'TGTG'-formula,
\be E_0=\frac{1}{2\pi}\int_0^\infty d\xi \, \Tr\ln(1-\M),
\label{2.3}\ee
and  $\xi$ is the imaginary frequency, $\om=i\xi$. The kernel $\M$ is the product of two other kernels,
\be \M(\vec{x},\vec{x}')=\int d{x''}
                            \N_A(\vec{x},\vec{x}'')\N_B(\vec{x}'',\vec{x}'),
\label{2.4}\ee
each of which  is given by
\be \N_{A,B}(\vec{x},\vec{x}')=
        \int d{x''}
                            G_0(\vec{x},\vec{x}'')T_{A,B}(\vec{x}'',\vec{x}'),
\label{2.5}\ee
in terms of the $T$-operators of the corresponding lattice. Further, the free Green's function,
\bea G_0(\vec{x}-\vec{x}')
    &=&\int\frac{d^3k}{(2\pi)^3}\,\frac{e^{i\vec{k}(\vec{x}-\vec{x}')}}{\om^2-\k^2-k_3^2+i0},
\label{2.6}\eea
enters this formula. We will use below also the following two representations,
\be\label{2.7}G_0(\vec{x})=\frac{e^{i\om|\vec{x}|}}{4\pi |\vec{x}|}
   =  \int\frac{d^2\k}{(2\pi)^2}\,\frac{e^{i\k(\x-\x')+i\Gamma |\x_3-x_3'|}}{2i\Gamma(k)},
\ee
with $\Gamma(k)=\sqrt{\om^2-k^2+i0}$ and $k=|\k|$.

The $T$-operators can be defined from
\bea&&     G(\vec{x},\vec{x}')\\\nn
&&=G_0(\vec{x}-\vec{x}')-\int dz\,\int dz'\,
        G_0(\vec{x}-\vec{z})T(\vec{z},\vec{z}')G_0(\vec{z}'-\vec{x}'),
\label{2.8}\eea
where $ G(\vec{x},\vec{x}')$ is the Green's function of equation \Ref{2.2} and can be related to the ansatz
\be   G(\vec{x},\vec{x}')=G_0(\vec{x}-\vec{x}')
        -\sum_{\n,\n'}        G_0(\vec{x}-\vec{a}_\n)\Phi^{-1}_{\n,\n'}G_0(\vec{a}_{\n'}-\vec{x}').
\label{2.9}\ee
Inserting this ansatz into the equation
\bea  &&\left(-\om^2-\Delta+g\sum_{\n}\left(
    \delta(\x-\vec{a}^{\rm \,A}_{\n})+\delta(\x-\vec{a}^{\rm \,B}_{\n})\right)
              \right)
    G(\vec{x},\vec{x}')\nn\\&&
    =\delta(\vec{x}-\vec{x}')
\label{2.10}\eea
and using the equation
\be  \left(-\om^2-\Delta \right)
    G_0(\vec{x},\vec{x}')=\delta(\vec{x}-\vec{x}')
\label{2.11}\ee
for the free Green's function, one comes to the equation
\be -\Phi^{-1}_{\n,\n'}+g \delta_{\n,\n'}
            -g\sum_{\m}G_0(\vec{a}_\n-\vec{a}_\m)\Phi^{-1}_{\m,\n'}  =0.
\label{2.12}\ee
With the definition
\be \Phi_{\n,\n'}=\frac{1}{g}\delta_{\n,\n'}-G_0(\vec{a}_\n-\vec{a}_{\n'})
\label{2.13}\ee
Eq. \Ref{2.12} becomes
\be \sum_\m \Phi _{\n,\m}\Phi^{-1}_{\m,\n'}=\delta_{\n,\n'}
\label{2.14}\ee
and $\Phi^{-1}_{\n,\n'}$ is the inverse matrix to $\Phi_{\n,\n'}$.

The diagonal elements of $\Phi_{\n,\n'}$ contain $G_0(0)$ which is not well defined. Referring to \cite{bord15-91-065027} for a discussion, we renormalize the coupling,
\be \frac{1}{g}-G_0(0) \to \frac{1}{g},
\label{2.15}\ee
and after that we get
\be \Phi_{\n,\n'}=\left\{\begin{array}{lc}\frac{1}{g},& (\n=\n')\\[5pt]
    -G_0(\vec{a}_\n-\vec{a}_{\n'}), &(\n\ne\n')\end{array} \right.
\label{2.16}\ee
which is well defined.

Now, comparing Eqs. \Ref{2.8} and \Ref{2.9}, we identify the $T$-operator to be
\be T^{\rm A}(\vec{z},\vec{z}')= \sum_{\n,\n'}\delta(\vec{z}-\vec{a}^{\rm \,A}_{\n})
    \Phi^{-1}_{\n,\n'}     \delta( \vec{a}^{\rm \,A}_{\n'}-\vec{z}')
\label{2.17}\ee
for one lattice and by the same formula with $B$ in place of $A$ for the other.
Inserting this expression into \Ref{2.5} we get
\be \N_A (\vec{x},\vec{x}')=
    \sum_{\n,\n'} G_0(\vec{x}-\vec{a}^{\rm \,A}_{\n})   \Phi^{-1}_{\n,\n'}
    \delta(\vec{a}^{\rm \,A}_{\n'}-\vec{x}')
\label{2.18}\ee
and, again, the same with $A\to B$. Further, inserting these into \Ref{2.4}, we get
\be     \M (\vec{x},\vec{x}')= \int dx''\, \N_A (\vec{x},\vec{x}'') \N_B (\vec{x}'',\vec{x}') ,
\label{2.19}\ee
for the kernel $\M$. Eqs. \Ref{2.18} and \Ref{2.19} represent the general expressions for the kernel entering the 'TGTG'-formula \Ref{2.3} for a generic lattice of delta functions.
In the following we consider first two dimensional lattices as given by \Ref{2.1}, and, subsequently one dimensional lattices (chains) as given by Eq. \Ref{2.1a}.
\subsection{Vacuum energy for two dimensional lattices}
In this subsection we consider lattices as given by Eq. \Ref{2.1}.
These have translational invariance with respect to a lattice step. Equivalently, we have $\a_\n-\a_{\n'}=\a_{\n-\n'}$, and the quantities entering Eq. \Ref{2.13} and \Ref{2.14}, depend only on differences, $\Phi_{\n,\n'}=\Phi_{\n-\n'}$ and the same for its inverse. As a consequence, the inversion of this matrix can be calculated by Fourier transform,
\be f_\n=a^2\int\frac{d\k}{(2\pi)^2} e^{i\k\a_\n}\tilde{f}(\k),\quad \tilde{f}(\k)=\sum_\n e^{-i\k\a_\n}f_\n.
\label{2.1.1}\ee
This way we get
\be \Phi^{-1}_{\n,\n'} =   a^2\int\frac{d\k}{(2\pi)^2} e^{i\k(\a_\n-\a_{\n'})}
        \frac{1}{\tilde{\phi}(\k)}
\label{2.1.2}\ee
with
\be \tilde{\phi}(\k) = \sum_\n e^{-i\k\a_\n} \Phi_\n.
\label{2.1.3}\ee
Here we insert \Ref{2.16} and come to
\bea \tilde{\phi}(\k) &=& \frac{1}{g}-{\sum_\n}' e^{-i\k\a_\n} G_0(\a_\n), \nn\\
                    &=& \frac{1}{g}-\frac{1}{4\pi}J_1(\om,\k),
\label{2.1.4}\eea
where Eq. \Ref{2.7} was used and we introduced
\be J_1(\om,\k)={\sum_\n}' \frac{1}{|\a_\n|}e^{i \om |\a_\n|+i \k \a_\n}.
\label{2.1.5}\ee
As usual, in the primed sum the term with $\n=0$ is dropped.
Finally, we mention the relation
\be \sum_\m  e^{i\k\a_\m } \Phi^{-1}_{\m,\m'} = \frac{1}{\tilde{\phi}(\k)}e^{i\k\a_{\m'} },
\label{2.1.6}\ee
which follows with the Fourier transform \Ref{2.1.1}.

Now we return to the kernel $\N$, \Ref{2.18}, and insert \Ref{2.7} for the Green's function,
\begin{widetext}
\be  \N_{A}(\vec{x},\vec{x}')=\sum_{\n,\m}
    \int\frac{d^2\k}{(2\pi)^2}\,\frac{e^{i\k(\x-\a_\n-\c)+i\Gamma |x_3-b|}}{2i\Gamma(\k)} \Phi^{-1}_{\n-\m}
                \delta(\a_\m+\c-\x')\delta(b-x_3').
\label{2.1.7}\ee
Next we use \Ref{2.1.6} and get
\bea  \N_{A}(\vec{x},\vec{x}')
    &=&  \int\frac{d^2\k}{(2\pi)^2}\,\frac{e^{i\k(\x-\c)+i\Gamma |x_3-b|}}{2i\Gamma(\k)\tfi(\k)}
    \sum_{\m} e^{-i\k \a_\m} \delta(\a_\m+\c-\x')\delta(b-x_3').
\label{2.1.9}\eea
\end{widetext}
The corresponding expression for $ \N_{B}(\vec{x},\vec{x}')$ follows from here with $\c=0$ and $b=0$. Using these, we can write down the kernel $\M(\vec{x},\vec{x}')$, Eq. \Ref{2.4},
\begin{widetext}%
\bea \M(\vec{x},\vec{x}')  &=&
    \int dx{''}
    \int\frac{d^2\k}{(2\pi)^2}\,\frac{e^{i\k(\x-\c)+i\Gamma |x_3-b|}}{2i\Gamma(\k)\tfi(\k)} \sum_{\m} e^{-i\k \a_\m} \delta(\a_\m+\c-\x'')\delta(b-x_3'')
    \nn\\&& \cdot
     \int\frac{d^2\k'}{(2\pi)^2}\,\frac{e^{i\k'\x''+i\Gamma' |x_3''|}}{2i\Gamma(\k')\tfi(\k')} \sum_{\m'} e^{-i\k' \a_{\m'}} \delta(\a_{\m'}-\x')\delta(x_3').
\label{2.1.10}\eea
\end{widetext}
Carrying out the integration over $\vec{x}''$ using the delta functions, we come to a sum over $\m$. Before doing this sum, we split the momenta $\k$ and $\k'$ into quasi momentum and integer part,
\be \k=\q+\frac{2\pi}{a}\Nn,\quad \k'=\q'+\frac{2\pi}{a}\Mm,
\label{2.1.11}\ee
where $\Nn$ and $\Mm$ are integer vectors like $\n$ in $\a_\n$. The integration becomes $\int d^2\k=\int d^2\q\sum_\Nn$ and the components of $\q =( q_{1} ,q_2)$ are restricted to
$-\frac{\pi}{a} \le {q}_{1,2}<\frac{\pi}{a}$.
Now the sum over $\m$ appearing in \Ref{2.1.10} gives
\be \sum_\m e^{-i(\k-\k')\a_\m}=\left(\frac{2\pi}{a}\right)^2\delta^{(2)}(\q-\q'),
\label{2.1.12}\ee
where the dependence on $\Nn$ and $\Mm$ dropped out. Then \Ref{2.1.10} turns into
\begin{widetext}
\bea \M(\vec{x},\vec{x}')  &=& \frac{1}{a^2}
    \int\frac{d^2\q}{(2\pi)^2}\,  \sum_\Nn    \frac{e^{i\x(\q+\frac{2\pi}{a} \Nn)-i\frac{2\pi}{a}\Nn\c+i\Gamma |x_3-b|}}{2i\Gamma(\k)\tfi(\k)}
    \sum_{\Mm}
    \frac{e^{i\frac{2\pi}{a}\Mm\c+i\Gamma' |b|}}{2i\Gamma(\k')\tfi(\k')}
    \sum_{\m'} e^{-i\q \a_{\m'}} \delta(\a_{\m'}-\x')\delta(x_3').
\label{2.1.13}
\eea
\end{widetext}
This expression for $\M(\vec{x},\vec{x}')$ must be inserted into the generalized Lifshitz formula \Ref{2.3} under the sign of the trace.
In doing so, the delta function in one factor $\M(\vec{x},\vec{x}')$  turns the $\x$ in the next factor into $\a_{\m'}$. Again, the summation over $\m'$ gives a delta function for the quasi momenta. In this way, the products of the factors $\M(\vec{x},\vec{x}')$ is diagonal in $q$ and we can define
\be h(\om,\q) = \frac{1}{a^2}\sum_\Nn\,
    \frac{e^{i \Gamma(\k)b+i\frac{2\pi}{a}\Nn\c}}{2i\Gamma(\k)\tfi(\k)}
\label{2.1.14}\ee
with $\k$ given by \Ref{2.1.11}, and we rewrite \Ref{2.1.13} in the form
\begin{widetext}
\be  \M(\vec{x},\vec{x}') = a^2\int\frac{d^2\q}{(2\pi)^2}\,e^{i\q\x} h(\om,\q)h(\om,\q)^*
    \sum_\m e^{-i\q\a_\m}\delta(\a_\m-\x')\delta(x_3').
\label{2.1.15}\ee
\end{widetext}
In this way, the factors $\M(\vec{x},\vec{x}')$ entering \Ref{2.8} become diagonal. The $\x$-integration in the trace is to be taken over one cell and it turns the sum into unity,
\be \int d\vec{x}\,e^{i\q\x}\sum_\m e^{-i\q\a_\m}\delta(\a_\m-\x)\delta(x_3) =1.
\label{2.1.16}\ee
As a consequence, for the vacuum energy per lattice cell we get the expression
\be E_0 =\frac{1}{2}\int_0^\infty \frac{d\xi}{\pi}\, a^2\int\frac{d^2\q}{(2\pi)^2} \ln\left(1-|h(i\xi,\q)|^2\right).
\label{2.1.17}\ee
This is the final general formula for the vacuum energy of two lattices as given by Eq. \Ref{2.1}. The numerical result for the vacuum energy for one cell of the lattices is presented at Fig.~\ref{fig1}.
\subsection{Vacuum energy for one dimensional lattices (chains)}
In this subsection we consider lattices as given by Eq. \Ref{2.1a}.
The calculations are mostly in parallel to the preceeding subsection. The Fourier transform is now one dimensional,
\be f_\n=a\int\frac{dk_1}{2\pi} e^{ik_1an}\tilde{f}(k_1),\quad \tilde{f}(k_1)=\sum_\n e^{-ik_1an}f_n,
\label{2.2.1}\ee
where we used for the momentum the notation $k_1$ since the lattice is parallel to the $x$-axis. Following Eqs. \Ref{2.1.2} and \Ref{2.1.3}, we get
\be \tilde{\phi}(k_1) = \sum_ne^{-ik_1an}\Phi_n,
\label{2.2.2}\ee
which after renormalization gives
\be \tilde{\phi}(k_1) =\frac{1}{g}-\frac{1}{4\pi}
    \sum_{n\ne 0} \frac{e^{i\om|n|+ik_1an}}{a|n|}.
\label{2.2.3}\ee
The sum is  one dimensional and can be done explicitly,
\be \tilde{\phi}(k_1) =\frac{1}{g}-\frac{1}{4\pi a}
   \ln\left(1+e^{2i\om a}-2 \cos(k_1 a) e^{i \om a }\right).
\label{2.2.4}\ee
By the way, the zeros of this function determine the zones in the Kronig-Penney model.

Now we consider the kernel $\N_A$, Eq. \Ref{2.5}, and insert the Green's function from \Ref{2.7} and the relation
\be \sum_n e^{ik_1 an}\Phi^{-1}_{m m'}=\frac{1}{\tilde{\phi}(k_1)}e^{-k_1 a m'}
\label{2.2.5}\ee
to get
\begin{widetext}
\bea  \N_{A}(\vec{x},\vec{x}')
    &=&  \int\frac{d^2\k}{(2\pi)^2}\,\frac{e^{ik_1(x_1-an-c)+i\Gamma |x_3-b|}}{2i\Gamma(\k)\tfi(k_1)}
    \sum_{m} e^{-ik_1 a m} \delta(a m+c-x_1')\delta(x_2')\delta(b-x_3').
\label{2.2.6}\eea
In parallel to \Ref{2.1.10}, but carrying out the integration over $x''$, we get further
\bea \M(\vec{x},\vec{x}')  &=&  \frac{1}{a}
      \int\frac{d^2\k}{(2\pi)^2}\,
      \frac{e^{ik_1(x_1-c)+ik_2x_2+i\Gamma |x_3-b|}}{2i\Gamma(\k)\tfi(k_1)}
      \sum_{m} e^{-ik_1 a m}
    \nn\\&& \cdot
     \int\frac{d^2\k'}{(2\pi)^2}\,
     \frac{e^{ik_1'(am+c)+i\Gamma' |x_3''|}}{2i\Gamma(\k')\tfi(k_1')}
     \sum_{m'} e^{-ik_1'am'} \delta(a m-x'_1)\delta(x_2')\delta(x_3').
\label{2.2.7}\eea
\end{widetext}
We introduce a splitting of the momenta like \Ref{2.1.11}, but now only for $k_1$,
\be k_1=q+\frac{2\pi}{a}N, \quad  k_1'=q+\frac{2\pi}{a}M,
\label{2.2.8}\ee
where $N$ and $M$ are integers, and come to
\begin{widetext}
\bea \M(\vec{x},\vec{x}')  &=& \frac{1}{a}
            \int\frac{dq}{2\pi}
            \int\frac{dk_2}{2\pi}\sum_N
            \frac{e^{i q(x_1-c)+i\frac{2\pi}{a}Nc+ik_2x_2+i\Gamma|x_3-b|}}
            {2i\Gamma(\k)\tfi(k_1)}
\nn\\&& \cdot
      \int\frac{dk'_2}{2\pi}\sum_M
     \frac{e^{iqc+i\frac{2\pi}{a}M c +i\Gamma' |b|}}{2i\Gamma(k')\tfi(k_1')}
          \sum_{m} e^{-iq a m}
          \delta(a m-x'_1)\delta(x_2')\delta(x_3').
\label{2.2.9}\eea
\end{widetext}
Following the same discussion concerning the trace as in the preceding subsection, we are lead to introduce the notation
\be h(\om,q) = \frac{1}{a}  \int\frac{dk_2}{2\pi}\sum_N
      \frac{e^{i\frac{2\pi}{a}Nc+i\Gamma |b|}}{2i\Gamma(\k)\tfi(k_1)}
\label{2.2.10}\ee
with
\be \Gamma=\sqrt{\om^2-k_1^2-k_2^2+i0},\quad k_1=q+\frac{2\pi}{a}N.
\label{2.2.11}\ee
Finally, we get the vacuum energy for one cell of the chains,
\be E_0 =\frac{1}{2}\int_0^\infty \frac{d\xi}{\pi}\,
    a\int\limits_{-\pi/a}^{\pi/a}\frac{d q}{2\pi}
        \ln\left(1-|h(i\xi, q)|^2\right).
\label{2.2.12}\ee
This is the final general formula for the vacuum energy of two lattices as given by Eq. \Ref{2.1a}.
\begin{figure}[t]
\includegraphics[width=8cm]{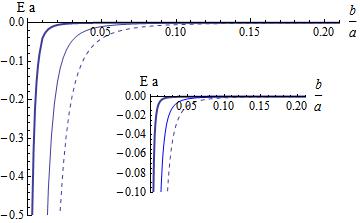}
\caption{The vacuum energy per unit as a function of distance between lattices  for different values of the coupling. Here lattices are exactly opposite to one another,  $c=0$. From left to right $g/a=0.01, 0.05,0.1$. Inset: the same for two parallel Dirac chains.}
\label{fig1}
\end{figure}

It should be mentioned that the formula for $h(i\xi,q)$ can be simplified since the integration over $k_2$ in \Ref{2.2.10} can be carried out. First, we perform the Wick rotation and define
\be \Gamma=i\gamma,\quad\gamma=\sqrt{\xi^2+k_1^2+k_2^2}.
\label{2.2.13}\ee
We get
\be h(i\xi,q) = \frac{1}{a}  \int\frac{dk_2}{2\pi}\sum_N
      \frac{e^{i\frac{2\pi}{a}Nc-\gamma b}}{-2\gamma\tfi(k_1)}.
\label{2.2.14}\ee
Changing the integration from $k_2$ for $\gamma$ and using the integral representation
\be K_0(z)=\int_0^\infty e^{z \cosh(\theta)} d\theta
\label{2.2.15}\ee
one comes to
\be h(i\xi,q) = \frac{1}{2\pi a}  \sum_N
      \frac{K_0(\sqrt{\xi^2+k_1^2}b)} {-2\tfi(k_1)}\, e^{i\frac{2\pi}{a}Nc}
\label{2.2.16}\ee
with $\tilde{\phi}(k_1)$ given by Eq. \Ref{2.2.4}.
%
The numerical result for the vacuum energy for one cell of the chains is presented at Fig.~\ref{fig1} (inset).

\begin{figure}[t]
\includegraphics[width=8cm]{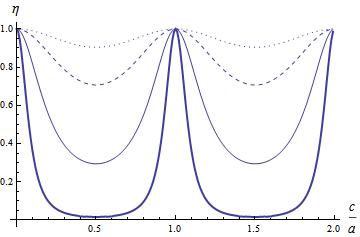}
\caption{The factor $\eta=E(c\ne0)/E(c=0)$ is plotted as a function of $c/a$ for $g/a=0.1$ and different separations $\beta=b/a$ of the chains. From bottom to top $\beta=0.1, 0.2, 0.4, 0.6$.}
\label{fig2}
\end{figure}

The influence of the displacement $c$ on the interaction between the chains is shown in Fig.~\ref{fig2}. Similar profiles may be found in \cite{buhm16-31-1641029} for the Casimir-Polder potential between an atom and a one-dimensional grating.

\section{Limiting cases}
In this section we discuss the limiting cases of the interaction lattices of delta functions.
There are two limiting cases for each, two and one dimensional lattices. One limiting case is for large separation, or equivalently for dimensional reasons, for small lattice step. The other is for small separation, or equivalently, large lattice step.

For large separation, the two dimensional lattices turn into planes, carrying a delta function potential. For these the Casimir effect was first investigated in \cite{bord92-25-4483}. The one dimensional lattices (chains) turn into lines, carrying a (two dimensional) delta function potential. This case was not considered so far. However, it turns out that that it is equivalent to the finite size cylinders with Dirichlet boundary conditions for turning their radius to zero, as considered in \cite{emig06-96-080403} and \cite{bord06-73-125018}.

For small separation, in the limit, only one delta function for the one lattice interacts with the closest one in the other lattice. This is equivalent to the Casimir-Polder interaction of two points carrying a delta function potential each. This is the scalar version of the well known interaction of two dipoles, which was considered within the scattering approach in \cite{powe93-48-4761}. Also, it can be considered as limiting case of the interaction of two spheres carrying delta function potential for vanishing radius of the spheres. The corresponding calculation is shown in the Appendix.

\begin{widetext}
The considered cases and their interrelations are shown in  Fig. \ref{fig3}.
\begin{figure}
\includegraphics[width=17.6cm]{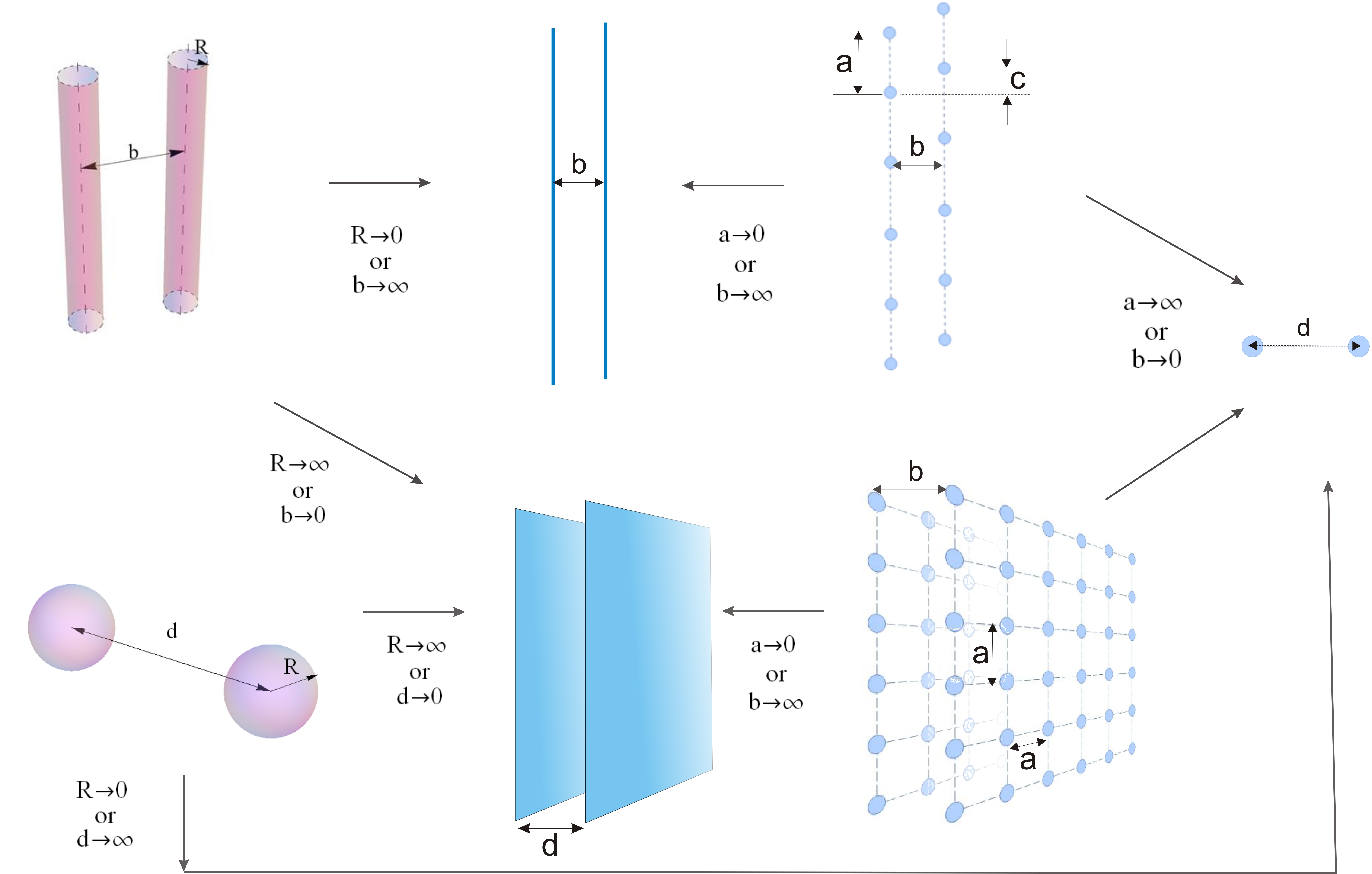}
\caption{Possible transitions from one dimensional lattices (chains) and two dimensional lattices of delta functions for large and small separation. For convenience, the two dimensional lattices are shown for $c=0$, i.e., without displacement. For large separation, the lattices turn into continuous plates and the chains into wires. For small separation both turn into the closes pints.
Also shown are the transitions from spheres and cylinders to plates and wires.}
\label{fig3}
\end{figure}
\end{widetext}

\subsection{Two dimensional lattices at large separation}
We start from Eq. \Ref{2.1.17} for the vacuum energy and consider the limit of lattice spacing $a\to0$, or equivalently, separation $b\to\infty$. At the end, we will get tow parallel continuous sheets for which we need the energy density per unit area which is $\frac{1}{a^2}E_0$. The integration region in $\q$, given by $-\frac{\pi}{a} \le {q}_{1,2}<\frac{\pi}{a}$, turns into a whole $\mathbb{R}^2$. The function $h(\om,\k)$, \Ref{2.1.14}, reads with imaginary frequency,
\be h(i\xi,\q) = \frac{1}{a^2}\sum_\Nn\,
    \frac{e^{- \gamma(\k)b+i\frac{2\pi}{a}\Nn\c}}{-2\gamma(\k)\tfi(\k)}
\label{3.1.1}\ee
with $\gamma=\sqrt{\xi^2+\k^2}$, $k_i=q_i+\frac{2\pi}{a}N_i$ (i=1,2). For $a\to 0$ all contributions except that for $\N=0$ vanish and we get $k_i=q_i$. Also we need the function $\tfi$, given by \Ref{2.1.4} and the function $J_1$, \Ref{2.1.5}, which reads now
\be J_1(i\xi,\k)={\sum_\n}' \frac{1}{|\a_\n|}e^{-\xi |\a_\n|+i \k \a_\n}.
\label{3.1.2}\ee
For $a\to0$, the sum turns over into an integral, which can be calculated easily,
\be J_1(i\xi,k)\raisebox{-5pt}{${\to \atop a\to 0}$} \ \frac{2\pi}{a^2\gamma},
\label{3.1.3}\ee
see Eq. (134) in \cite{bord15-91-065027}. This way we get
\be h(i\xi,k)\raisebox{-5pt}{${\to \atop a\to 0}$} \ r\,e^{-\xi b},
\label{3.1.4}\ee
where
\be r=\frac{1}{1-\frac{a^2}{g}2\gamma}.
\label{3.1.5}\ee
The vacuum energy \Ref{2.1.17} becomes
\be \frac{1}{a^2}E_0   \raisebox{-5pt}{${\to \atop a\to 0}$} \
    \frac12\int_0^\infty\frac{d\xi}{\pi}\int\frac{d\q}{(2\pi)^2}
    \ \ln\left(1-\left( r\,e^{-\gamma b}\right)^2\right).
\label{3.1.6}\ee
In fact, $r$, Eq. \Ref{3.1.5}, is the well known reflection coefficient for a plane carrying a delta function potential and \Ref{3.1.6} gives the Casimir effect for two such sheets.
The ratio $g/a^2$, having the meaning of coupling per unit area,  is the strength of the delta function potential in the corresponding equation, where the delta function is one dimensional and well defined such that this strength carries over from the equation to the reflection coefficient without change. This is in opposite to the strength $g$ in Eq. \Ref{2.1.17} which after renormalization \Ref{2.15} has little to do with the $g$ in equation \Ref{2.2}. The other way round, one can use the established above relation to the continuous sheets as normalization of the coupling $g$ after renormalization.
\subsection{One dimensional lattices (chains) at large separation}
This case is to a large extend in parallel to the preceding subsection. We start from \Ref{2.2.12} and consider the energy per unit length, $\frac{1}{a}E_0$. Further, we substitute $\xi\to\xi/b$ and $\q\to q/b$. This way we get a factor $1/b^2$ in front and all further dependence is on $a/b$.

Again in the function $h(i\xi,k_1)$, Eq. \Ref{2.2.4},  accounting for \Ref{2.2.8}, only $N=0$ gives a non vanishing contribution. A somehow different feature appears from $\tfi$, \Ref{2.2.17}, which has a logarithmic behaviour now,
\begin{widetext}
\bea \tfi(k_1)    &=&
        \frac{1}{g}-\frac{b}{4\pi a}\left(-\frac{a}{b}\xi+\ln\left(2\left(\cosh\left(\frac{a}{b}\xi\right)^2-\cos\left(\frac{a}{b} k_1\right)^2\right)\right)\right) ,   \nn\\
       &=&
                  \frac{1}{g}-\frac{b}{4\pi a}
       \left(2\ln\left(\frac{a}{b}\right)-\frac{a}{b}\xi+\ln\left(\xi^2+k_1^2\right)+\dots\right) ,\nn \\
        &\raisebox{-5pt}{${\to \atop a\to 0}$} & -\frac{\ln(\frac{a}{b})}{2\pi a/b}.
\label{3.2.1}\eea
\end{widetext}
The last line is the leading order and one observes that the dependence on the coupling $g$ is lost. Inserting into \Ref{2.2.16} we get
\be h(\xi,k_1)   \raisebox{-5pt}{${\to \atop a\to 0}$}  \
        \frac{b}{\ln(a/b)}\,K_0\left(\sqrt{\xi^2+k_1^2}\right).
\label{3.2.2}\ee
Finally, inserting into \Ref{2.2.12}, we get
\begin{widetext}
\be \frac{1}{a}E_0 \raisebox{-5pt}{${\to \atop a\to 0}$} \
        \frac{1}{2b^2} \int_0^\infty\frac{d\xi}{\pi}\int\frac{dk_1}{2\pi}
    \ \ln\left(1-\left(\frac{1}{\ln(a/b)} K_0\left(\sqrt{\xi^2+k_1^2}\right)\right)^2\right),
\label{3.2.3}\ee
\end{widetext}
and further expanding the logarithm,
\be \frac{1}{a}E_0 \raisebox{-5pt}{${\to \atop a\to 0}$} \
        -\frac{1}{8\pi b^2} \frac{1}{\ln(a/b)^2},
\label{3.2.4}\ee
where $\int_0^\infty dp\,p\,K_0(p)^2=\frac12$ was used.

It is to be mentioned, that this result coincides with the large separation limit of the interaction between two parallel cylindrical shells carrying   $\delta$-potentials (see Appendix B),  considered also in \cite{milt08-41-155402}.  The same limit may be
reproduced from~\cite{bord06-73-125018}, Eq. (88), for the interaction of a conducting cylinder with a conducting plane after taking into account the different geometries and a factor of 2 for the polarizations.
\subsection{Two dimensional lattices at short separation}
Here we consider the vacuum energy \Ref{2.1.17} for the interaction of two dimensional lattices for small separation, or equivalently for large lattice steps. In this case the integration region for the integration over $\q$ in \Ref{2.1.17} shrinks to zero and this integral gives
\be a^2 \int\frac{d^2\q}{(2\pi)^2} f(\q)\raisebox{-5pt}{${\to \atop a\to \infty}$} \
        f(0),
\label{3.3.1}\ee
where $f(\q)$ is some arbitrary function. In the function $h(\om,\q)$, \Ref{2.1.4}, we first consider the function $J_1(\om,\k)$, \Ref{2.1.5}. For $a\to\infty$, it simply vanishes. As a consequence, $\tfi(\k)$, \Ref{2.1.4}, turns into
\be \tfi(\k) \raisebox{-5pt}{${\to \atop a\to \infty}$} \ \frac{1}{g}.
\label{3.3.2}\ee
Now in the function $h(\om,\q)$, the sum over $\N$ turns into an integration according to
\be \frac{2\pi}{a}\N\raisebox{-5pt}{${\to \atop a\to \infty}$} \\k,\qquad
        \frac{1}{a^2}\sum_\N
        \raisebox{-5pt}{${\to \atop a\to \infty}$} \
        \int\frac{d\k}{(2\pi)^2}.
\label{3.3.3}\ee
This way, we get
\be h(i\xi,\q) \raisebox{-5pt}{${\to \atop a\to \infty}$} \
        \int\frac{d\k}{(2\pi)^2}   \frac{g}{-2\gamma}e^{-\xi b+i\k\c}.
\label{3.3.4}\ee
The integration can be carried out and after some calculation we arrive at
\be h(i\xi,\q) \raisebox{-5pt}{${\to \atop a\to \infty}$} \
         \frac{g}{4\pi d}e^{-\xi d},
\label{3.3.5}\ee
where $d=\sqrt{b^2+c^2}$ is the separation between the closes delta functions from the two lattices. With these, the vacuum energy turns into
\be E_0\raisebox{-5pt}{${\to \atop a\to \infty}$} \
    \frac12\int_0^\infty\frac{d\xi}{\pi}\ln\left(1-\left(\frac{g}{4\pi d}e^{-\xi d}\right)^2\right),
\label{3.3.6}\ee
which is the Casimir-Polder interaction of two points at separation $d$ carrying a delta function potential each. It coincides with the interaction of two spheres carrying delta function potential at large separation (or small radius). The corresponding formula is displayed in the Appendix. Again, we remark that the coupling $g$ in \Ref{3.3.6} is after renormalization and, therefore, is not fixed.
\subsection{One dimensional lattices (chains) at short separation}
We consider the vacuum interaction of two chains as given by Eq. \Ref{2.2.12} for short separation. Similar to the preceding subsection the integration over q shrinks to a point,
\be a \int\frac{dq}{2\pi} f(q)\raisebox{-5pt}{${\to \atop a\to \infty}$} \
        f(0),
\label{3.4.1}\ee
Next we consider the function $\tfi(k_1)$, \Ref{2.2.4}, and have
\be \tfi(k_1) \raisebox{-5pt}{${\to \atop a\to \infty}$} \ \frac{1}{g}.
\label{3.4.2}\ee
in the function
$h(\om,\q)$. The sum over $\N$ turns into an integration according to
\be \frac{2\pi}{a}N\raisebox{-5pt}{${\to \atop a\to \infty}$} \k,\qquad
        \frac{1}{a}\sum_N
        \raisebox{-5pt}{${\to \atop a\to \infty}$} \
        \int\frac{dk}{2\pi}.
\label{3.4.3}\ee
and we get from \Ref{2.2.16}
\bea h(i\xi,q) &\raisebox{-5pt}{${\to \atop a\to \infty}$} &
        g\int_{-\infty}^\infty\frac{dk}{2\pi}  K_0(\sqrt{\xi^2+k^2}b)e^{ikc},\nn\\
        &&=  \frac{g}{4\pi d}e^{-\xi d},
\label{3.4.4}\eea
where again the integration over $k$ was carried out and $d=\sqrt{b^2+c^2}$ is the separation between the closest delta functions from the two chains.

Now, inserting \Ref{3.4.1} and \Ref{3.4.4} into \Ref{2.2.12}, we get
\be E_0\raisebox{-5pt}{${\to \atop a\to \infty}$} \
    \frac12\int_0^\infty\frac{d\xi}{\pi}\ln\left(1-\left(\frac{g}{4\pi d}e^{-\xi d}\right)^2\right),
\label{3.4.5}\ee
which is the same expression as \Ref{3.3.6} and represents the Casimir-Polder interaction of two points carrying delta function potential as discussed at the end of the preceding subsection.
\section{Comparison with pairwise summation}
A special feature of van der Waals and Casimir forces is their multi particle character. While in certain cases a pairwise summation may give a good approximation, in general it will not. In this section we compare the results of the exact calculation for point scatterers with the result of pairwise summation, i.e., without multi particle forces.
\begin{figure}[t]
\includegraphics[width=8cm]{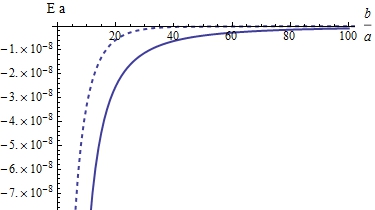}

\caption{The difference between exact result \Ref{2.2.12} (dashed line) and pairwise summation of individual Casimir-Polder interactions for chains \Ref{3.4.5}, $g/a=0.1$.
}
\label{fig4}
\end{figure}

To this end we perform a pairwise summation of all individual Casimir-Polder interactions
\be F(z)=\sum_{\mathbf{n}}F_n \cos(\varphi_n), \;  F_n(z)=-\left.\frac{dE_{CP}(r)}{dr}\right|_{r=r_n},
\ee
where $E_{CP}$ is given by \Ref{3.3.6}.
For two chains $\mathbf{n}=n$, $r_n=\sqrt{z^2+n^2 a^2}$ and $\cos(\varphi_n)=1/\sqrt{1+(an/z)^2}$; for two 2D-lattices a double summation is required $\mathbf{n}=\{n_1,n_2\}$, $r_\mathbf{n}=\sqrt{z^2+n_1^2 a^2+n_2^2 a^2}$ and $\cos(\varphi_n)=1/\sqrt{1+(an_1/z)^2+(an_2/z)^2}$.
And finally, the interaction energy per  one $\delta$-function is
\be
E_{pw}(b)=\int_{b}^{\infty}F(z)dz=\sum_{\mathbf{n}=-\infty}^{\infty}E_{CP}(r_\mathbf{n})
\ee

As follows from Section III, Eqs. \Ref{3.3.6} and \Ref{3.4.5}, the pairwise summation is a good approximation at short distances, where the
exact formulas tend to a two-point Casimir-Polder interaction.
At medium and large distances the  pairwise summation overestimates the vacuum energy.
The corresponding curves for Dirac chains at medium distances are presented at  Fig. \Ref{fig4}.
Here the pairwise summation of Casimir-Polder interactions is performed for $N=1000$ Dirac $\delta$-s.
The evaluation according to the exact formula, Eq. \Ref{2.2.12}, was also truncated at $N=1000$ in Eq. \Ref{2.2.4}, though the result only weakly depends on $N$ at medium and large distances.

The result of pairwise summation depends considerable on the coupling $g$ at large separations, while the asymptote of the exact result, Eq. \Ref{3.2.4}, is coupling independent.

\section{Conclusions}\label{s7}
We considered $T$-operators for a two-dimensional and a one dimensional lattices of $\delta$-functions. These  involve    lattice sums, which
can be expressed in terms Hurwitz zeta function. Further we used this  $T$-operator to formulate the kernel in the TGTG formula for the dispersion interaction  of two such lattices. This can be viewed as a kind of generalized Lifshitz formula and a represents a finite (converging) expression for the interaction energy.  We considered the cases of the interaction of two parallel two-dimensional lattices and,   of 1-dimensional parallel lattices (chains). The generalization to rotations is left for future work.

We consider in detail to limiting cases and show the transition from lattices to plans for large separation and to the Casimir-Polder interaction of two lattice sites at small separation. These limiting cases  are in agreement with earlier results. Our formulas appear to interpolate between these and establish the link between these..

\section*{acknowledgement}
We acknowledge partial support from the Heisenberg-Landau Programme.

\appendix

\section{The Casimir effect for two spheres carrying delta function potentials}
In this appendix we display  basic formulas for the interaction of two spheres carrying delta function potential ('semitransparent' spheres) using the by now well known scattering approach ('TGTG'-formula). Although such kind of calculations are, in much more general form, contained in a number of papers, for example in \cite{emig08-08-04007}, specific formulas for one of the most simple special cases may be of use. The first calculation involving delta function potential is \cite{bord07-75-065003}, which was focused on the corrections beyond PFA.

The basic setup is given by the equation
\begin{widetext}
\be  \left(-\om^2-\Delta+\frac{g}{4 \pi R^2}
        \left(\delta(|\vec{x}|-R)+\delta(|\vec{x}-\vec{d}|-R)\right)
              \right)\phi(\vec{x})=0,
\label{A.1}\ee
\end{widetext}
where $R$ is the radius of the spheres, one at the origin, the other at separation $d=|\vec{d}|$. These delta functions are one dimensional and the coupling $g$ does not undergo any renormalization.

Within the chosen approach, we need for the $T$-operator for a single sphere. In this case the equation is
\be  \left(-\om^2-\Delta+\frac{g}{4 \pi R^2}
        \delta(|\vec{x}|-R)              \right)\phi(\vec{x})=0.
\label{A.2}\ee
The scattering problem for a single sphere was considered, for example, in \cite{bord99-59-085011} and we use some notations from there.  The Green's function for Eq. \Ref{A.2}, in a spherical basis, is
\be G(\vec{x},\vec{x}')=\sum_{lm}Y_{lm}(\Om)d_l(r,r')Y_{lm}(\Om')^*
\label{A.3}\ee
and the free radial Green's function is
\be d^{(0)}(r,r')=\frac{1}{rr'}j_l(\om r_{<})h^{(1)}_l(\om r_{>}),
\label{A.4}\ee
where $j_l(z)=\sqrt{\pi/(2z)}J_{l+1/2}(x)$ and $h^{(1,2)}_l(z)=\sqrt{\pi/(2z)}H^{(1,2)}_{l+1/2}(x)$ are spherical Bessel functions. Their Wronskian is $j_l'(z)h_l(z)-j_l(z)h_l'(z)=1/iz^2$. The delta function in \Ref{A.2} results in matching conditions on the radial function $d_l(r,r')$, which must be continuous and obey
\be \pa_r d_l(r,r')_|{_{r=R+0}}-\pa_r d_l(r,r')_|{_{r=R-0}}=-\frac{g}{4 \pi R^2}d_l(R,r').
\label{A.5}\ee
With the ansatz
\be d_l(r,r')=d^{(0)}(r,r')-d^{(0)}(r,R)\Phi^{-1}(\om)d^{(0)}(R,r')
\label{A.6}\ee
we get
\be \Phi^{-1}(\om)=\frac{i\om g/4\pi}{1+\frac{g}{4\pi R}i\om j_l(\om R)h_l(\om R)}.
\label{A.7}\ee
After Wick rotation, this expression becomes
\be \Phi^{-1}(i\xi) =\frac{-g R/4\pi}{1+\frac{g}{4\pi R^2}I_{l+\frac12}(\xi R)K_{l+\frac12}(\xi R)}
\label{A.8}\ee
with the modified Bessel functions $I_\nu(z)$ and $K_\nu(z)$. Eq. \Ref{A.8} is in agreement with Eq. (23) in \cite{bord99-59-085011}.

The 'TGTG'-formula can be written in a spherical basis,
\be E_0=\frac12\int_0^\infty\frac{d\xi}{\pi}\Tr \ln(1-\M)
\label{A.9}\ee
where the kernel is
\be \M_{l,l';m}=\sum_{l''=|l-l'|}^{ l+l'}  \N_{l,l'';m}\N_{l'',l';m}
\label{A.10}\ee
and the trace is over the orbital momenta. All expressions are diagonal in the magnetic quantum number $m$. The kernels in \Ref{A.10} are
\bea \N_{l,l';m} ~~~~~~~~~~~~~~~~~~~~~~~~~~~~~&& \\ \nn
=\frac{1}{R^3}\sqrt{\frac{\pi}{2\xi R}}K_{l''+\frac12}(\xi d)H^{l''}_{l,l'}
     &&   I_{l+\frac12}(\xi R) I_{l'+\frac12}(\xi R)\Phi^{-1}(i\xi)
\label{A.11}\eea
with the factors $H^{l''}_{l,l'}$ resulting from the   matrix elements for the transition between the two centers. These are given, e.g., by Eq. (10.129) in \book~ or Eq. (5) in
\cite{Teol11-84-125037}, and correspond to the $\cal U$ in \cite{emig08-08-04007}, Eq. (2.25). For $g\to\infty$, this formula turns into that for Dirichlet boundary conditions on the spheres.

For us, the important property is that for $d\to\infty$, the leading order comes from $l=l'=l''=0$ and with $H^0_{0,0}=1$ we get
\be \N_{00}\raisebox{-5pt}{${\to \atop d\to \infty}$} \ \frac{g}{4\pi d}E^{-\xi d}.
\label{A.12}\ee
The dependence on the radius $R$ of the spheres drops out in this limit. By inserting this into \Ref{A.9} we get
\be E_0\raisebox{-5pt}{${\to \atop d\to \infty}$} \
    \frac12\int_0^\infty\frac{d\xi}{\pi}\ln\left(1-\left(\frac{g}{4\pi d}e^{-\xi d}\right)^2\right)
\label{A.13}\ee
for the interaction of two spheres carrying delta function potential in the limit of large separation.

\section{The Casimir effect for two cylinders carrying delta function potentials}

The derivation of  the vacuum energy for two parallel cylinders carrying delta function potentials may be found, for example in \cite{milt08-41-155402} or \cite{milt08-77-045005},
\be
\frac{E}{L}=\frac{1}{4\pi}\int\limits_0^{\infty} d\xi \, \xi \, \Tr \ln(1-\M),
\label{B.1}
\ee
with $\M$ given by (\ref{A.10}) and
\bea \N_{l,l';m}=\frac{g R K_{l+l'}(\xi d) I_{l'}^2(\xi R)}{1+g R I_{l'}(\xi R) K_{l'}(\xi R) }.
\label{B.2}\eea
Here we compute the large distance limit of this vacuum energy. The leading contribution for large separation $d$ comes from the s-wave. After the substitution  $\xi \to \xi/d$ in (\ref{B.1}), the relevant matrix elements entering the trace may be rewritten in the form
\be
\M_{ll'}=\delta_{l}^{0}\delta_{l}^{0}\frac{g^2 R^2 K_0^2(\xi) I_0^4(\xi R/d)}{(1+g R I_0(\xi R/d) K_0(\xi R/d))^2}.
\label{B.3}
\ee
With allowance for the behaviour of the Bessel functions at small arguments,
$I_0(x)=1+O(z^2)$ and $K_0(x)=-\gamma-\ln(x/2)+O(z^2)$, one arrives at the expression
\bea
&&\M_{ll'}=\\
&&\frac{\delta_{l}^{0}\delta_{l'}^{0}K_0^2(\xi)}{\ln^2(R/d)}\left[1+\frac{\gamma+\ln(\xi/2)-(gR)^{-1})}{\ln(R/d)} \right]^{-2} +O((R/d)^2), \nn
\label{B.4}
\eea
which can be easily expanded in powers of small $1/\ln(R/d)$.
This expansion we substitute into \Ref{B.1} and  in the leading order  arrive at
\be
\frac{E}{L}=-\frac{1}{4\pi d^2}\int\limits_0^{\infty} d\xi \; \xi \frac{K_0^2(\xi)}{\ln^2(R/d)}=-\frac{1}{8\pi d^2 \left(\ln(R/d)\right)^2},
\label{B.5}
\ee
which coincides with \Ref{3.4.5} which was obtained from the chains.

%

\end{document}